# NANOMECHANICS OF ADVANCED POLYMER FIBRES

Ph. Colomban

Laboratoire de Dynamique, Interaction et Réactivité
UMR 7075 (CNRS-Université Pierre et Marie Curie)
2 rue Henry-Dunant
94320 Thiais, France
philippe.colomban@glvt-cnrs.fr

**ABSTRACT**

The micro/nano-structural evolution before and after tensile/compressive loading, fatigue and ultimately, failure has been studied by Raman (and IR) microspectroscopy for PBO, PET, PA66, PP, silk and hair using three probes: low wavenumber collective modes at <150cm$^{-1}$ as representative of the crystalline/ordered and amorphous chains, stretching and bending modes, as representative of the C-C/ C-N polymeric backbone, and localized vibrations (OH, NH) to probe the inter-macromolecule segment distance. Wavenumber and bandwidth distribution across fibre diameters reveal different types of skin/core heterogeneity. The *in situ* analysis at different strain/pressure levels shows that amorphous chains in the fibre accommodate the stress differently. The *post mortem* analysis shows that amorphous domains can be higly stressed during the failure and remnant stress can be measured.

## 1. INTRODUCTION

The challenge for the nanotechnologies is to achieve perfect control on nanoscale-related properties. This obviously requires correlating the parameters of the synthesis process with the resulting nanostructure and analysing the changes at the nanoscale, *in situ*, at work. The behaviour of fibres in compression and tension is not well known. It is not only the case for new systems like, for example, the PBO Zylon$^{TM}$ (paraphenylenebenzonitrozyl) but also for older materials like polyamide fibres or high performance natural fibres (*Bombyx mori* or spider silks, hairs). Fibre strength is a factor of primary importance in composite strength, because the resistance of single filaments and multifilament tows are required for prediction of ultimate failure of composites. Fatigue failure is also a key point in many applications. Actually even single filaments have a complex micro/nanostructure and can be considered as composite materials. Complex processes of extrusion-extension or spinning from a liquid state crystal regulate the texture of high performance synthetic and natural fibres. The mechanisms correlating texture and nano-/micro-properties remain to be clarified.

Not all conventional techniques are suited for the analysis of nanosized/nanostructured materials but IR/Raman spectroscopies have already proven to be one of them [1-4]. Two kinds of parameters will influence the vibrational signatures : i) those acting on vibration "mechanics" like atomic mass, bond strength or the system geometry will set peaks wavenumbers (the eigenfrequencies of matter vibration), ii) those acting on the "charge transfer" (iono-covalency, band structure,…) will set peak intensity. Raman bandwidths are characteristic of the local order, more specifically the short range arrangement in the first (0.1-0.5nm) and second (~0.5-5 nm) atomic shells. Thus, Raman spectroscopy, and for some extent IR spectroscopy, analyse specifically the crystalline and amorphous components of a materials. Furthermore, on the basis of chemical bonds anharmonicity, any-stress induced interatomic distance alteration changes the atomic vibration wavenumbers and offers a tool to study deformation and mechanics at the nanometric scale. We will try here to summarize the broad possibilities that Raman (and to some extend IR) microscopes offer in the study of the texture and the (nano)mechanical behaviour of fibres by closely analyzing the vibrational spectra in the whole wavenumber range. Additionally, the techniques of deuteration developed for the study of compounds with hydrogen bonding [4] will be considered. We selected synthetic fibres exhibiting different properties, PBO, an elastic high modulus material, polyamide 66 (PA66) and poly(ethyleneterphthalate) (PET), two engineering thermoplastic polymers, isotactic polypropylene (i-PP) a viscoelastic materials and natural fibres (silks and hair), both with a viscoelastic behaviour. Stress-strain curves are compared in Fig. 1. Note that the ultimate stress is directly correlated to the density, i.e. to the number of chemical bond in the unit volume which depends from the fibre micro/nanostructure. This review examines and compares the behaviour of the different types of fibres under tensile (and fatigue) loading until fracture and under hydrostatic pressure. Emphasis is given to the discussion between the fibre (nano)structure, structural heterogeneity and macromechanical properties.

## 2. EXPERIMENTAL

Raman and IR vibrational analysis was performed on single fibres extracted from the yarn as described in refs [1-3]. Single fibres were mounted in the grips and were loaded in a Universal Fibre Tester [1] as shown in Fig. 2, and the fibres were submitted to different strain levels up to their fracture and their Raman spectra *in situ* recorded. In addition, some fibres, which had been broken during fatigue tests, have been analysed at points ahead of the fatigue crack tip [3].

A single fibre small fragment (length < 150 µm) was cut and placed in the diamond-anvil cell. A ruby ball was added for the pressure measurement (Fig. 2). The diamonds used where selected for their low fluorescence, allowing "good" transparency, both for IR and Raman measurements.

A "XY" spectrograph (Dilor, Lille, France) equipped with a double monochromator filter and a back-illuminated, liquid nitrogen-cooled, CCD detector was used to record Raman spectra between 10 and 2000 $cm^{-1}$, using 514.5 and 647.1 nm exciting lines ($Ar^+$-$Kr^+$ laser). The power of illumination ranged between 0.5 mW and 10 mW. Backscattering illumination and collection of the scattered light were made through an Olympus confocal microscope (long working distance Olympus x50, x80 or x100 objectives, total magnification x500, x800 or x1000).

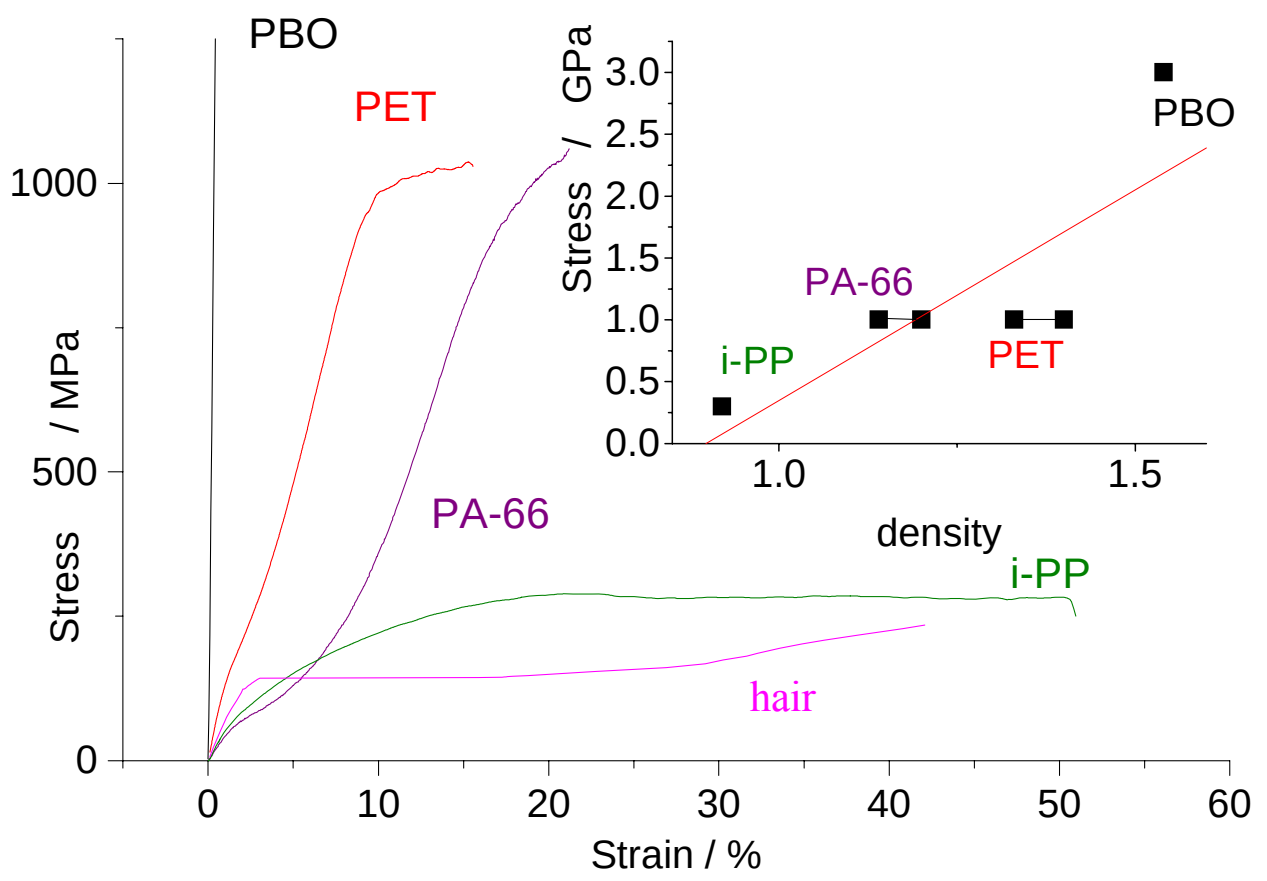

Figure 1 : Stress-strain plots for studied fibres (note PBO stress reaches ~3 GPa) and correlation between ultimate strength and density.

The infra-red microscope/ATR is an Equinox 55 Fourier Transform Michelson interferometer (Bruker, France) with a ATR "golden gate" input designed to examine the surface (a few microns) of a material brought into "perfect" contact with the ATR crystal. A "Irscope II" microscope with Cassegrain objectives makes it also possible to work in transmission with the same diamond anvil cell used for Raman measurements. The spot diameter ~200 µm is reduced to ~20 µm with a diaphragm, to separately analyse the core and skin of a fibre.

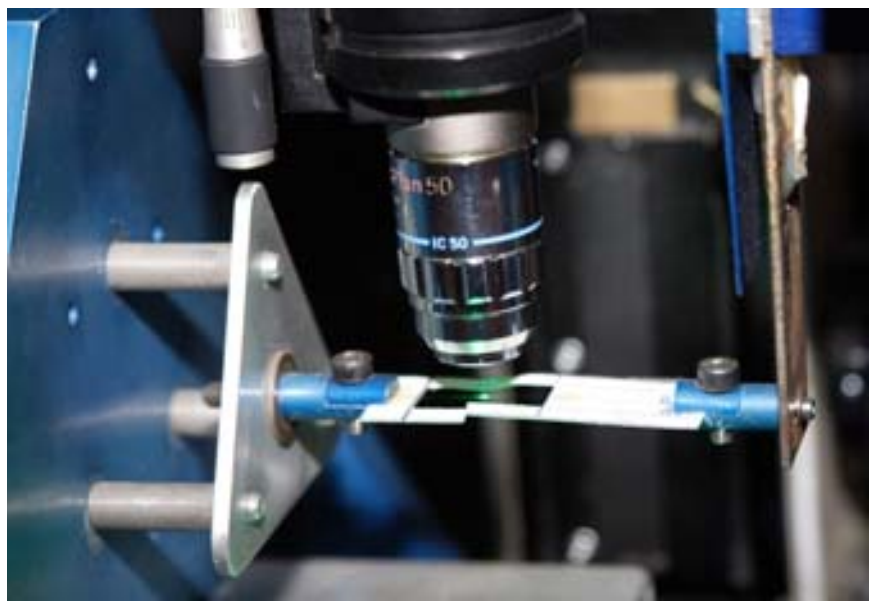

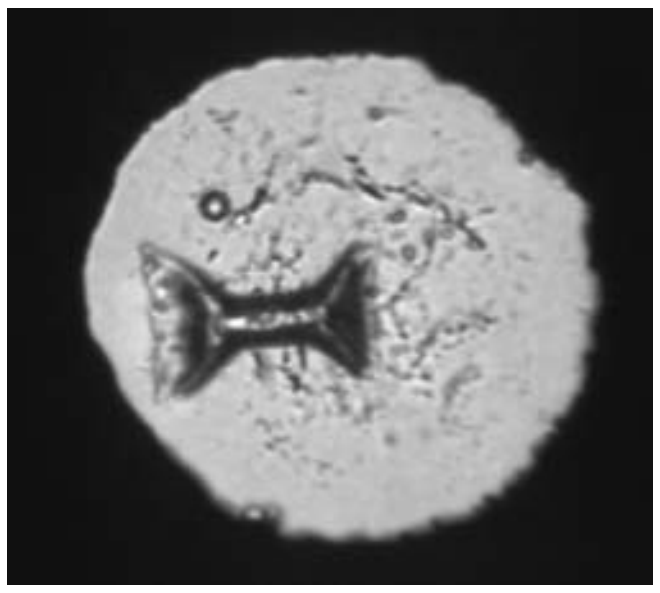

Figure 2 : Detail of the experimental setting for Raman examination of the single filament fiber under controlled load; PET fibre in the diamond-anvil cell (cell diameter ~200µm).

## 3. CRYSTALLINE/AMORPHOUS CONFORMATIONS

Not only the position (wavenumber) but also the form (Lorentzian or Gaussian), the width, and the polarized modes of the Raman and IR

components give valuable information on the composition (elongation modes), the structure (polarizations and collective low energy modes) and the texture (polarization mappings) of the fibres. The comparisons of polarized signatures for // (H) and ⊥ (V) configurations obtained by changing the orientation of the fibre with respect to the spectrometer slit entrance (Fig. 3), is a good mean of measuring the axial character of the fibre and to compare the crystallinity and texture of different fibres, including their precursors [2]. The Raman cross section of the O-H vibrators being very weak, a discussion on the water content has to rely on the IR spectra obtained with the Irscope (Fig. 4), in particular using the band at ~3450 $cm^{-1}$ that strongly weakens after drying. In the case of the X-H vibrators, the intrinsic mode width (resulting from the structural disorder alone) is obtained only in the case of an isotopic dilution [H]/[D] <10% : very light vibrators normally couple in a complex way (mechanical, electric, quantum) with the other phonons and only isotopic dilution, by removing the essence of these couplings, it becomes possible to obtain a spectrum giving details on the vibrators environment [4].

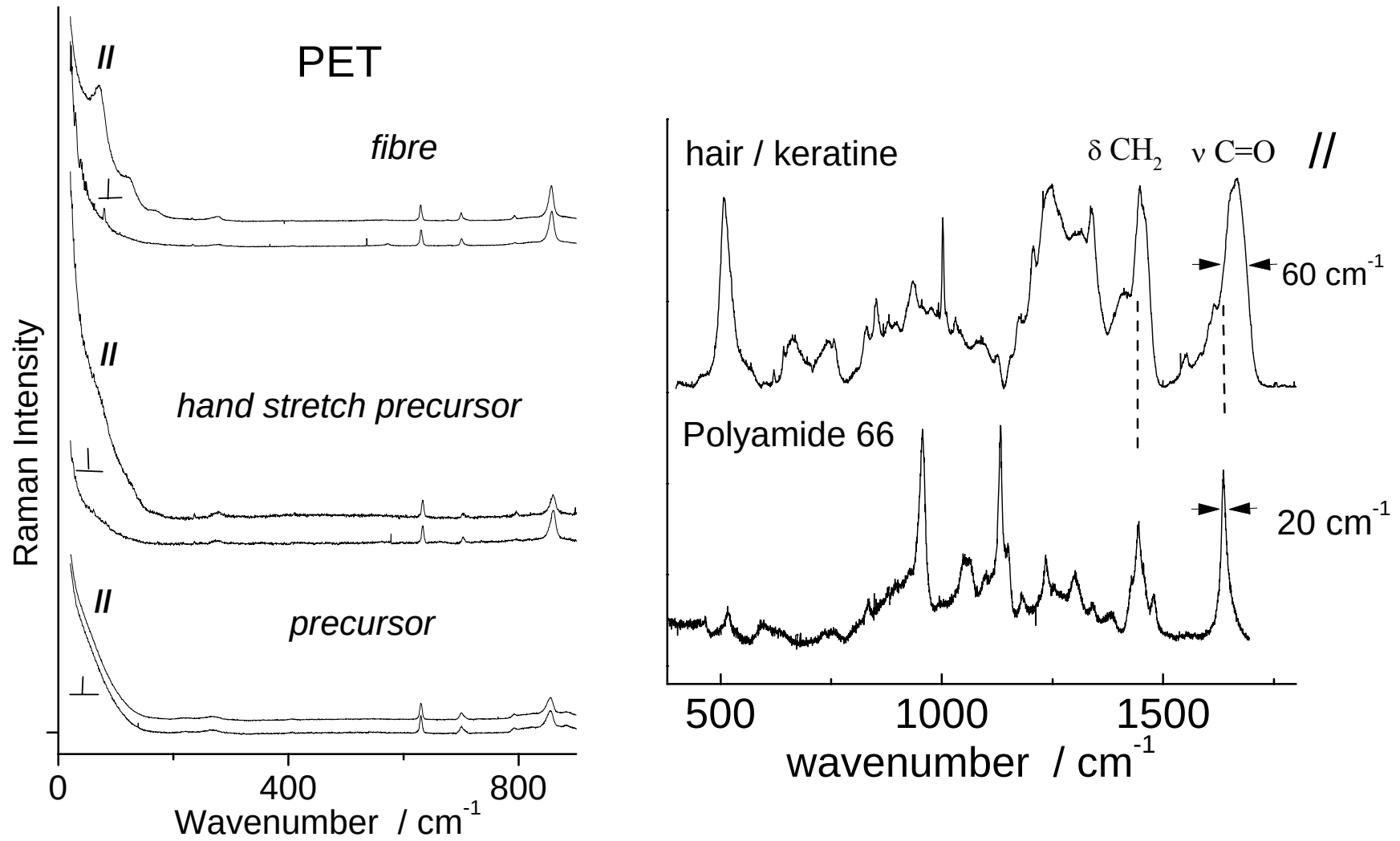

Figure 3 : Examples of polarised Raman spectra showing the (PET) crystallisation and texturation of a amorphous precursor filament under tension into a high crystalline oriented fibre; comparison between polarised spectra of high crystalline polyamide and low crystalline keratine fibres.

High resolution instruments now make it possible to analyse the low wavenumber region, where modes have a collective character. This is obvious from the comparison of polarized spectra (Figs 3-4): the modes at about 80-150 $cm^{-1}$ assigned to modes involving the movement of the chains of the unit-cell exhibit the stronger degree of polarization. Roughly speaking, these mode can be described as a translation (T') of the macromolecule with respect to adjacent ones. These modes involve modifications of the electron-richer bonds and hence gives rise to very strong peaks. The greater the number of atoms involved in the motion, the lower the wavenumber will be. On the other hand, inter-chain coupling will shift the peak toward high wavenumbers. These low wavenumber collective modes may be preferred for the study of the relationship between the chain structure and the macroscopic mechanical behaviour. Differentiation between crystalline/ordered macromolecules (with a polarised Lorentzian Raman signature) and amorphous/disordered (down-shifted Gaussian un-polarised homologue) informs on the fibre nanostructure.

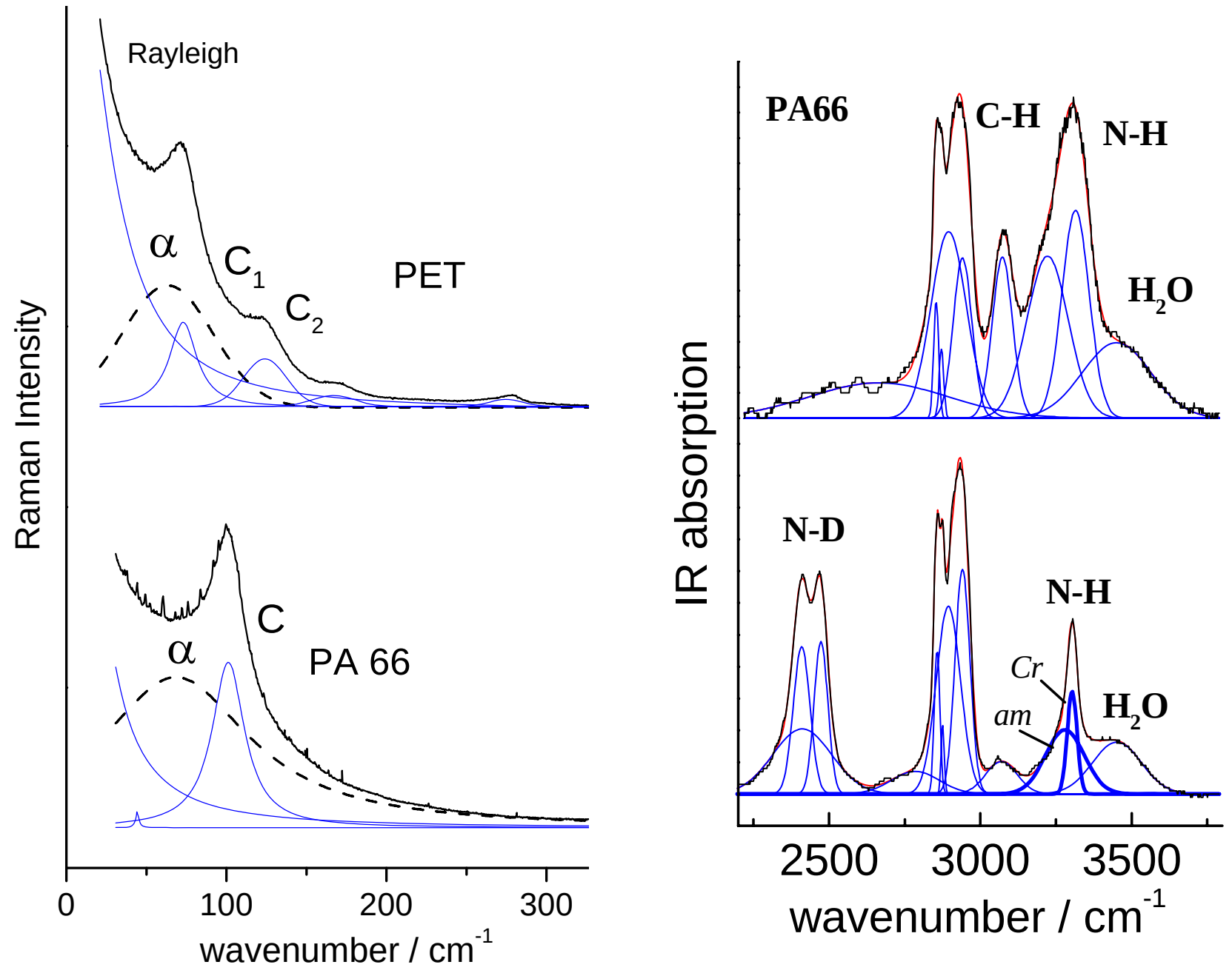

Figure 4 : Collective low wavenumber Raman modes in // polarisation for PET and PA 66 single fibres. Comparison between IR signatures of pristine and deuterated (H isotopically diluted) PA 66 single fibre. Broad and narrow components respectively show amorphous (α,am) and crystalline (C,Cr) conformations.

## 4. CORE/SKIN TEXTURES

The recording of series of spectra across the fibre diameter (e.g. Fig. 5) makes it possible to analyse the texture anisotropy of the various fibres [1-3]. A core-skin effect is obvious from the wavenumber shift across the fibre diameter. For instance in PA 66 fibre, variations indicate a weak tension of the crystalline chains and a strong compression of the amorphous bonds in the fibre core. The strong width reduction of the "amorphous" macromolecular chain component in the fibre core undoubtedly indicates a much better organization, at the local scale, of the amorphous "phase". This characteristic, related to the slower cooling of the fibre core disappears by annealing above $T_g$. PBO fibre diameter line scan shows rather similar behaviour: the bandwidth of the amorphous chain signature strongly increases in the fibre centre. On the contrary the crystallinity of studied PET fibres does not vary across the fibre diameter and a wavenumber shift down is observed for from the periphery to the fibre center indicating tension of the macromolecules.

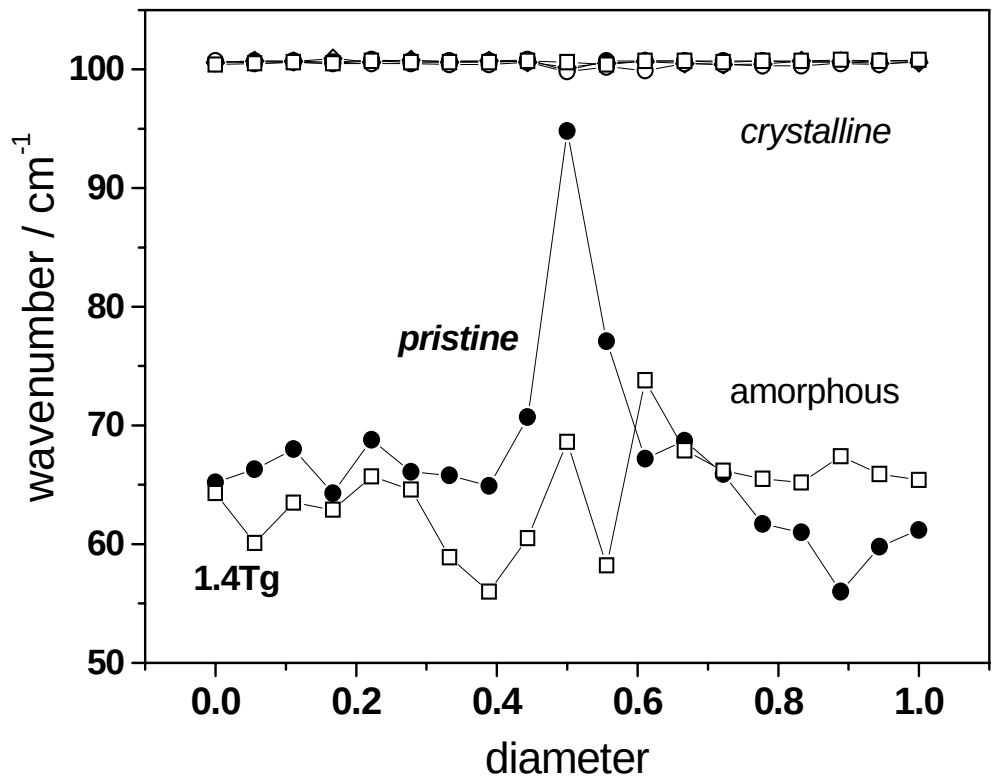

Figure 5 : Plot of the crystalline and amorphous low wavenumber component across the diameter (~30µm) of a PA 66 single fibre, before and after heating above the Tg temperature.

## 5. THE RELATIONSHIP BETWEEN NANOMECHANICS AND MACROMECHANICS

The vibrational spectroscopy through chemical bond nano-scale probe allows to distinguish the changes in conformations (nanostructure), chain orientation, the increase of intra and inter-macromolecule distances and the strain/stress of the chemical bonds. Very different behaviours are observed for crystalline and amorphous signatures. As for the macro stress/strain curve recorded using the Universal Testing Tester two or more important stages in the mechanisms of deformation can be seen in Figs 6-7 for the two examples considered, PET and keratine fibres (see [1-3] for PA 66, PP and PBO). The initial plateau behaviour, observed for both crystalline and amorphous chains signature (Fig. 4), is consistent with an initial alignment of the chains without energy dissipation up to a threshold which differs as a function of the molecular chain (PET ~300 MPa (4%), PP ~50 MPa (5% of strain), PA 66 ~300-550 MPa (10-12%)) and also as a function of the fibre grade and producer). Above this threshold two opposite behaviours are observed for the signature of crystalline chains : i) a very small increase in wavenumber is observed for PP and PET (followed by a plateau regime), indicating that the fibre stretch leads to a slight compression of the crystalline chains; ii) a small down-shift is observed for PA 66 and keratine fibres. These two opposite behaviours provide information about the structural arrangement. Note the perfect plateau due to α/β transition in keratine fibre takes place after an elastic regime. The analysis of PA 66 fibres broken in fatigue highlighted a state of compressive stress of the amorphous phase close to the point of initiation of the rupture and its progressive decrease over 200 to 300 µm beyond this point, all occurring as if the rupture in fatigue resulted from the loss of viscoelasticity, at certain points, of the amorphous phase.

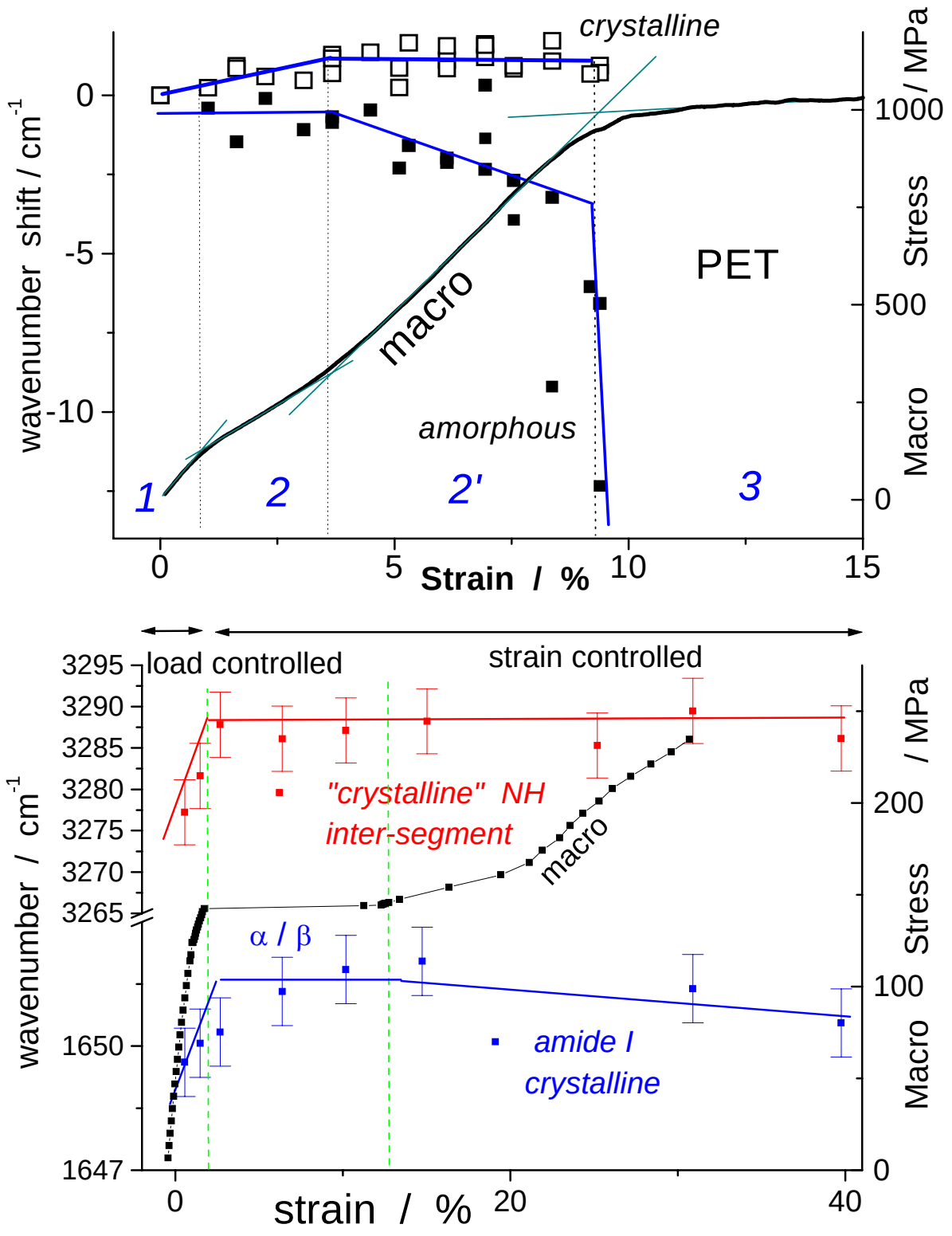

Figure 7 : Comparison between the nano- and macromechanic behaviour for PET (low wavenumber probe) and (dry) keratine (νNH and amide I probes) fibres; measurements have been made under controlled load (PET, hair <4%) or strain (hair >4%).

The comparison of the macroscopic (stress/strain plot) and nanometric (wavenumber shift versus applied strain) shows that stage threshold observed at the (macromolecule) nano and macroscopic scale almost coincides.

Rather similar conclusion can be made from data obtained using the diamond anvil cell. Under hydrostatic pressure the coming together of the chains (reduction in $\nu_{N-H}$) can be clearly seen and the existence of a threshold from which the geometry of the PA 66 fibres is modified. On the other hand, the compression of PBO macromolecule starts immediately, according to the very high Young modulus resulting from the very rigid aromatic cycles.

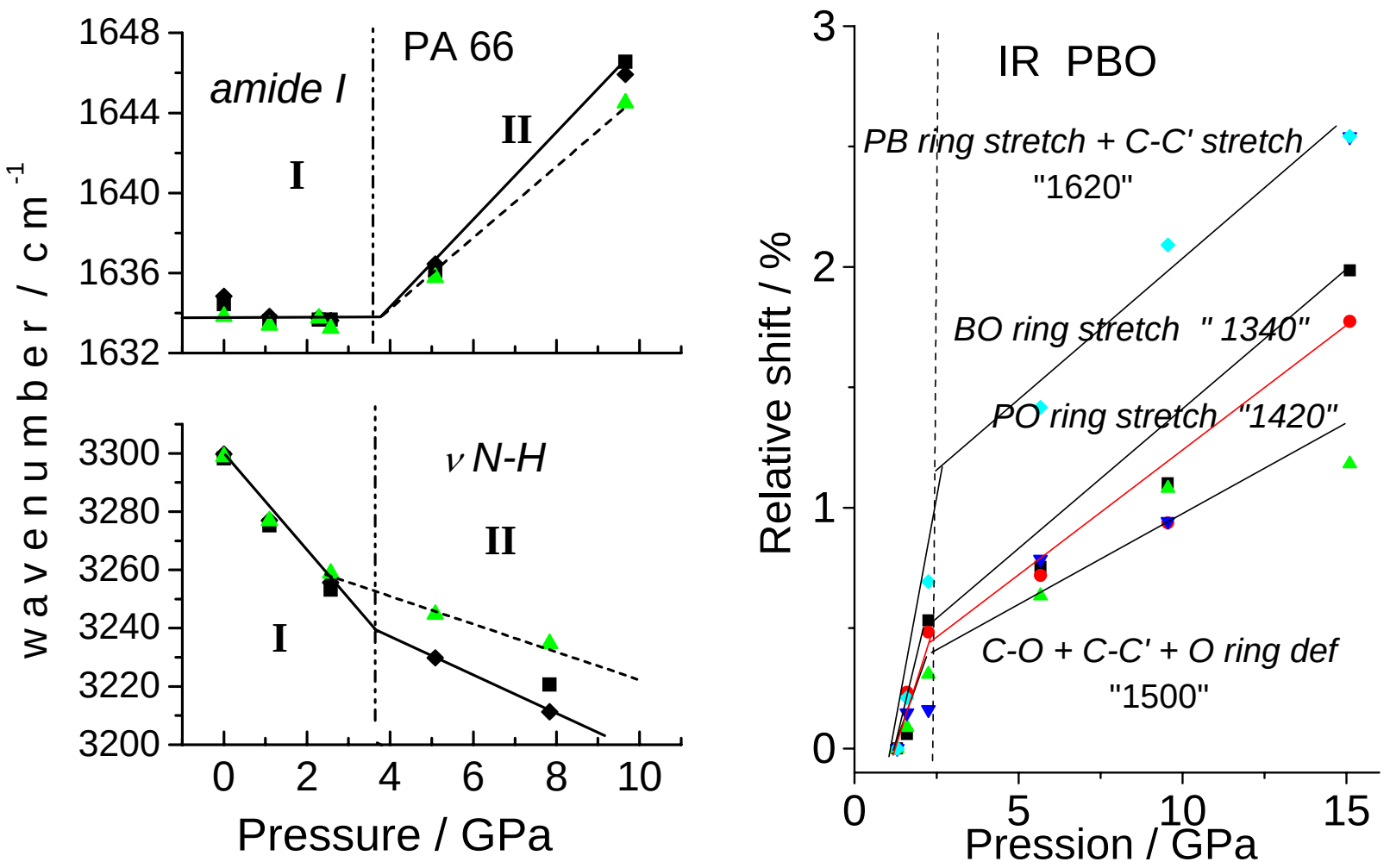

Figure 8 : Example of Raman and IR wavenumber shift as a function of hydrostatic pressure for PA 66 and PBO single fibre for different probes. Note the PA 66 macromolecule accomodates the pressure below 4 GPa.

## 6. CONCLUSION

The possibility of separately analysing "crystalline" and "amorphous" macromolecule conformation/structure allows a better comprehension of the processes of (nano)rupture and fatigue of any fibre built with convalent-bonded molecular entities giving a good vibrational signature.

## REFERENCES

1- **Marcellan A., Colomban Ph., Bunsell A.**, "(Nano)structure, skin/core and tension behaviour of polyamide fibres", *J. Raman Spectrosc.*, **35** (2004) 308-15.

2- **Colomban, Ph., Herrera Ramirez J.M., Paquin R., Marcellan A., Bunsell.,** "Micro-Raman study of the fatigue and fracture behaviour of single Pa66 fibres : Comparison with single PET and PP fibres", *Engn. Fract. Mech.,* **73** (2006) 2463-75.

3- **Colomban Ph., Sagon G., Lesage M. Herrera Ramirez J.M.**., "Micro-Raman study of the compressive behaviour of advanced PA 66 polyamide fibres in a diamond-anvil cell," *Vib. Spectrosc.*, 3**7** (12005) 83-90.

4- **Colomban Ph**., "Micro-analyse Raman et Ir de fibres à haute performance en traction et compression », in Benzeggagh M.L. & Lamon J. editors, Matériaux Composites, Comptes-rendus des 14èmes Journées Nationales sur les Composites (JNC14), Compiègnes 22-24 mars 2005, vol. 1, AMAC, Paris, ISBN 2-95051176-7, 3-12.